# Extrapolating Zernike Moments to Predict Future Optical Wave-fronts in Adaptive Optics Using Real Time Data Mining

Akondi Vyas[#,*], M B Roopashree[*], B Raghavendra Prasad[#]

[#]*Laser Lab, CREST, Indian Institute of Astrophysics, Shidlaghatta Road, Hosakote, India*
$^1$vyas@iiap.res.in
$^2$roopashree@iiap.res.in
$^3$brp@iiap.res.in

[*]*Indian Institute of Science, Malleswaram, Bangalore, India*

*Abstract*— We present the details of predicting atmospheric turbulence by mining Zernike moment data obtained from simulations as well as experiments. Temporally correlated optical wave-fronts were simulated such that they followed Kolmogorov phase statistics. The wave-fronts reconstructed either by modal or zonal methods can be represented in terms of Zernike moments. The servo lag error in adaptive optics is minimized by predicting Zernike moments in the near future by using the data from the immediate past. It is shown statistically that the prediction accuracy depends on the number of past phase screens used for prediction and servo lag time scales. The algorithm is optimized in terms of these parameters for real time and efficient operation of the adaptive optics system. On an average, we report more than 3% improvement in the wave-front compensation after prediction. This analysis helps in optimizing the design parameters for sensing and correction in closed loop adaptive optics systems.

## I. Introduction

Adaptive Optics (AO) is an imaging technology that improves the vision quality by adaptively correcting the distortions in the optical path of the source [1]. AO has shown significant improvement in the image quality with large telescopes like the Keck and Gemini telescopes [2, 3]. An AO system comprises of three major components:
- Wave-front sensing instrument that determines the shape of the incoming wave-front,
- Wave-front corrector, generally a deformable mirror that compensates the optical distortions by imposing a conjugate wave-front over the distorted one,
- Control algorithm calculates the command values to be addressed to the deformable mirror from sensor measurements.

The residual wave-front error in AO systems includes the wave-front reconstructor errors, servo lag errors, and errors due to CCD noise. The main reason for servo lag errors in adaptive optics is the low light level of astronomical objects which forces the astronomers to have large exposure time scales. Hence the sensed wave-front is a cumulative effect over the exposure timescales and in addition gets corrected only after the exposure time. In other words, the wave-front that is sensed is not corrected immediately, but after a delay equal to the servo lag error. Within this time the wave-front undergoes changes due to turbulence dynamics.

It is possible to minimize the errors in wavefront compensation due to servo lag errors by predicting the structure of atmospherically distorted wavefronts that are arriving at a future instance [4]. Predictive optimal estimators have been developed earlier to incorporate the spatio-temporal statistics of atmospheric turbulence [5]. Linear prediction algorithms have been implemented in the case of zonal as well as modal wave-front reconstruction algorithms [6,7]. Artificial Neural networks proved to be an effective tool in the prediction of atmospheric turbulence based on frozen in turbulence approximation [8-10]. Some algorithms were also tested experimentally [11]. The prediction of high and low spatial frequency aberrations was made possible by using multigrid methods [12]. Fourier prediction methods were adopted for faster implementation [13]. The frozen in turbulence approximation was validated by long duration temporal wave-front correlation experiments [14,15]. The algorithms have been optimized for arbitrary closed loop delays [16].

Data mining is a knowledge recovery technology from large and real-time data sets [17]. Since the temporally evolving turbulence is a time series data, prediction of the future can be made by applying suitable mining on the data parameters. Modally as well as zonally reconstructed wave-fronts can be represented in terms of the Zernike polynomial basis set by computing the complex Zernike moments. Zernike moments are computed for all the wave-fronts in the time series as suggested by Hosny [18]. Using the Zernike moments of the immediate past, Zernike moments of the future are predicted by extrapolating through regression analysis. The real and imaginary parts of the image moments are predicted independently. The dynamic behaviour of turbulence has a direct effect on the servo lag. It is also important to optimize the number of data points required to predict the future.

The servo lag errors in adaptive optics may arise either due to the bandwidth constraints set by individual components of the AO system or some other external factors like the finite exposure time. This is even more critical in astronomical case where incident light intensity is low. With the development of large telescopes AO and Multi Conjugate AO (MCAO) technologies also saw a dramatic leap [19]. For sensing, either a natural guide star or an artificially placed laser guide star is a requirement. Even at good sites the minimum exposure time required is ~5msec and 1msec for a natural and laser guide stars respectively. The instrumental time lag between sensing and correction, which includes the read out speed of the CCD; control algorithm delay; and delays in addressing the correcting element by the control unit, can be of the order of 1msec [9]. Hence, effectively there is a delay in compensation of the wave-front by 2-6msec. This timescale is within the decorrelation time which is ~17msec for sites like Hanle (India) [20]. In other words, the wave-fronts recorded within 17msec will be well correlated making the future prediction of turbulent wave-fronts a possibility.

The second section details the theory behind the prediction process using Zernike moments of temporally evolving wave-fronts using data mining. The third section gives the computational results of the improvement in the performance of the AO system after the prediction process. In the last section, the conclusions are presented.

## II. Data Mining Using Zernike Moments

The projections of the three dimensional wave-fronts form two dimensional images called phase screens. Simulation of temporally evolving phase screens that satisfy Kolmogorov spatial statistics is described in the first subsection [21]. This subsection also contains a description of the experimental procedure to simulate the effect of a linearly evolving turbulence. The simulated phase screens are arranged in a time series and are represented in terms of Zernike polynomials.

Zernike polynomials are a complete set of orthogonal polynomials that can be defined over a unit disk. Any two dimensional image function can be reconstructed using Zernike polynomials by calculating suitable coefficients that weigh each of the Zernike polynomials called Zernike moments of the image function. Each of the phase screens $p(x, y)$ can hence be represented in terms of Zernike polynomials $Z(x, y)$,

$$p(x, y) = \Sigma\, a^i\, Z^i(x, y) \quad (1)$$

Here, $a^i$'s represent Zernike moments. The details of the computation of Zernike moments of two dimensional image functions as suggested by Hosny [18] are sketched in the second subsection.

To extract useful information out of the calculated Zernike moments it is important to stack them together in the order of the time evolution of the phase screens. The third subsection details the formation of a large data matrix from the calculated Zernike moments.

Piecewise linear representation is the most generally used approximation of a time series data, which involves two major steps, segmentation and linear fitting. The fourth subsection describes the segmentation methodology and fitting algorithms used. The last subsection gives the steps to be followed to predict the future phase screens from the Zernike moment data.

### A. Simulation of temporally evolving phase screens

We implemented two methods to simulate temporally evolving phase screens. The first method is based on mathematical modelling of atmospheric turbulence. The second method is experimental where an evolving spatial turbulence effect was simulated by placing a moving aberrator in front of a wave-front sensor.

Noll established a relationship between the correlation matrix of Zernike moments and the spatial statistics of atmospheric turbulence following Kolmogorov statistics [22]. This Zernike representation of the Kolmogorov spectrum was used to compute Zernike moments, $a^i$'s such that the spatial statistics of turbulence are satisfied [23]. To simulate temporally evolving atmospheric turbulence, a simple wind model suggested earlier was put to use [24]. This simple wind model assumes that the velocity contribution comes from wind translation and local convection. The simulated phase screens are arranged in a time series fashion $\{A_1, A_2, ..., A_N\}$, where N is the number of simulated phase screens. The correlation of the $j^{th}$ phase screen with $A_1$ reduces with increasing j, as shown in Fig. 1. Each point on the time series represents two dimensional image functions. Many data sets were formed by simulating many such time series.

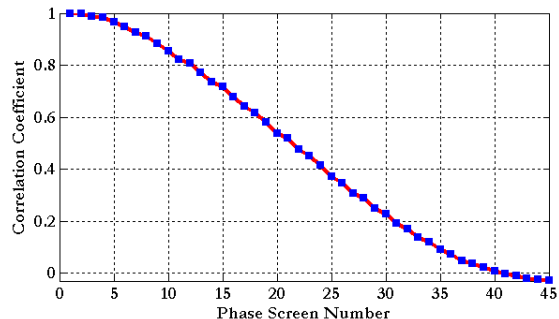

Fig. 1. The correlation drops with increasing time-simulated phase screens

The experimental setup to obtain evolving turbulence is shown in Fig. 2. Laser light is spatially filtered using the five axis Newport spatial filter, S.F and then collimated using a 20 cm focal length lens L1. A compact disk case (CDC) was used as an aberrator in the path of the collimated beam that falls on the Spatial Light Modulator (SLM) based Shack Hartmann Sensor (SHS). A SHS is a wave-front sensing device that closely reconstructs the aberration introduced in the optical path from the measurement of the local gradients. The CDC closely follows Kolmogorov spatial statistics. The focal plane of the lenslet array was reimaged on the CCD using a lens, L2 of focal length 15cm.

Images of the focal plane of the SHS were captured by shifting the aberrator in steps of 10μm along the plane perpendicular to the propagation of the light. The phase screens were reconstructed from these images by estimating phase at discrete points from the average local slope measurements of SHS [25]. The reconstructed phase screens were then arranged in a time series {$A_1$, $A_2$, ..., $A_N$}. The correlation between the first phase screen with the later screens drops as shown in Fig 3.

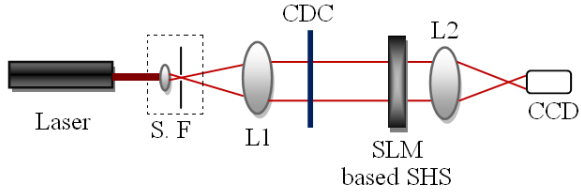

Fig. 2. Experimental setup to obtain evolving turbulence

For simplicity N was fixed to be 350 in both cases. A sample set of evolving phase screens are shown in Fig. 4. In both the cases, correlation drops below 25% after 25 phase screens. Since any two randomly generated phase screens are correlated by 25%, we can safely assume that if the correlation drops below this value, the phase screens are decorrelated. Since $25^{th}$ phase screen corresponds to the decorrelation time (17ms), a time lag of 2-6 ms corresponds to the phase screens numbered 3-9. The phase screen that needs to be predicted hence is a direct function of the servo lag error and the decorrelation time.

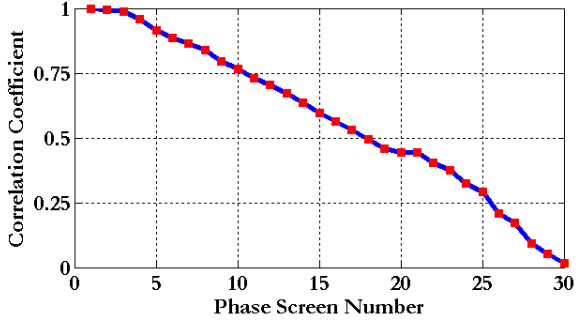

Fig. 3. Drop of correlation coefficient in the case of experimentally obtained data

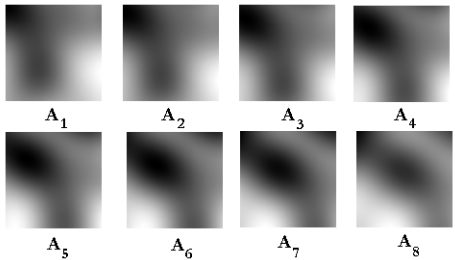

Fig. 4. Evolving Phase Screens

*B. Computation of Zernike Moments*

Zernike moments are computed as suggested by Hosny [18]. Since Zernike moments are mathematically complex, many recursive relations were developed by many authors for easy computation of Zernike moments [26]. These recursive methods lead to approximation errors. In the method adopted by Hosny, the approximation errors are removed by calculation of exact Zernike moments. The geometric errors are minimized by applying a proper image mapping. The computations are made faster by storing vectors and matrices that are constants and are repeatedly used.

A sample phase screen and the reconstructed phase screen using 40 orders of Zernike moments are shown in Fig. 5. The real and complex Zernike moments vary as shown in Fig. 6. The y-axis in the plot represents the Zernike moment value.

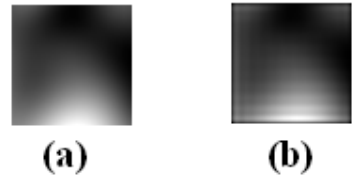

Fig. 5. (a) Sample Phase Screen (b) Phase Screen Reconstructed using Zernike moments

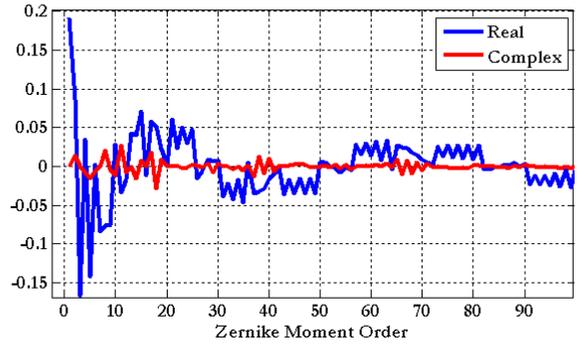

Fig. 6. Real and Complex parts of Zernike moments of the phase screen shown in Fig. 5 (a)

The advantage of extrapolating the Zernike moments for prediction is that Zernike polynomials represent the primary optical aberrations closely. Zernike polynomials can be alternatively represented in terms of their radial and azimuthal indices [n, m]. Zernike polynomial [1, 1] represents tilt in one direction and [2, 2] represents astigmatism. The evolution of the real and complex parts of Zernike moments corresponding to [1, 1] and [2, 2] is depicted in Fig. 7. It can be observed that the moments are slowly evolving making the possibility of mining more efficient. The evolution of Zernike moments will be smoother and slow only if the atmospheric turbulence is slowly evolving.

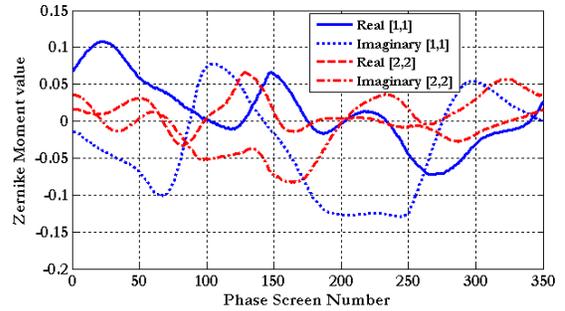

Fig. 7. Evolution of the Zernike moments (1,1) and (2,2)

*C. Formation of Data Matrices*

For the phase screen $A_j$, the Zernike moments are given by $a_j^i$ where, index 'i' represents the moment number and index 'j' represents the phase screen number in the time series. $a_j^i$ can be written as a sum of real ($r_j^i$) and imaginary quantity($c_j^i$), $a_j^i = r_j^i + c_j^i$. Since 'r' and 'c' are double indexed quantities, they can be represented using a single matrix of size N×M, where N number of phase screens are represented using M number of Zernike moments. In other words, the Zernike moments of the first phase screen are arranged in the first row of the data matrix as illustrated in Fig. 8.

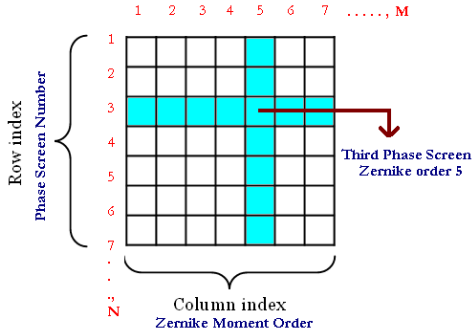

Fig. 8. Data matrix formation from the moment data

The real and imaginary parts of the Zernike moments of the time series are written and stored in two matrices R, C.

*D. Segmentation*

This linear approximation supports fast, nearly exact and concurrent mining. The problem is hence to segment the time series data into smaller pieces and then represent each of the segments by straight lines. There exist many segmentation algorithms in literature like top-down, bottom-up, sliding windows and other hybrid techniques. In this paper we implemented the simplest segmentation algorithm where only the last segment is selected with a known segment size or segment length. This segment was then used for prediction. The segment size is a constant in this case throughout the data linearization procedure, which need not be the case for above mentioned segmentation algorithms.

The segmentation parameters are the segment size and the fitting algorithm. The optimum segment size depends on the data distribution. The fitting algorithm can be either linear interpolation or linear regression. Another parameter external to mining and internal to adaptive optics system is the servo lag delay, which decides the phase screen that is to be predicted.

*E. Mining Zernike Moments*

In the process of predicting the future wave-fronts, it should be possible to predict the Zernike moments of the future phase screens by extrapolating the data of the Zernike moments of the phase screens in the immediate past. The mining and prediction of Zernike moments involves the following steps:
- As and when the wave-front is sensed, it must be decomposed into its corresponding Zernike moments.
- The computed Zernike moments should be placed in the last row of the data matrix showed in Fig. 8.
- Form a subset matrix from the large data matrix by selecting a number of future phase screens equal to the optimum segment size.
- Extrapolate each of the columns to predict future Zernike moments corresponding to a phase screen that is ahead by a time equal to the servo lag timescales.
- Compute the future phase screen by using the extrapolated Zernike moment values.

The correlation of the actual (future) wave-front with the predicted one was then calculated and compared with the correlation between the actual (future) wave-front with the wave-front behind in time equal to the servo lag.

### III. COMPUTATIONAL RESULTS

Monte Carlo simulations were performed to test the prediction algorithm. Small portions of the data matrix are read and prediction is performed. For example, if the segment size is fixed at 5, then we pick up the first 5 rows and M columns of the matrices R and C to form sub matrices $R_1$ and $C_1$. If the servo lag timescale corresponds to a phase screen number of 4, then $R_1$ and $C_1$ are extrapolated to compute the new Zernike moments at that instance. The future phase screen is then calculated by reconstruction using Zernike moments. The phase screen predicted in this fashion is then compared with actual (future) phase screen corresponding to the $9^{th}$ row of the data matrix. This process is repeated by shifting the position of the first phase screen in the segment formed from the data matrix.

Two important parameters were optimized in this study, the segment size and the servo time lag. Studying the servo time lag will help us to optimize the exposure time, which is an important part in the servo lag error.

*A. Segment Size*

The segment size was varied from 2 to 10. The phase screen number to be predicted was fixed at 5 corresponding to a time delay of less than 3ms. The effect of varying segment size for different data sets is plotted in Fig. 9.

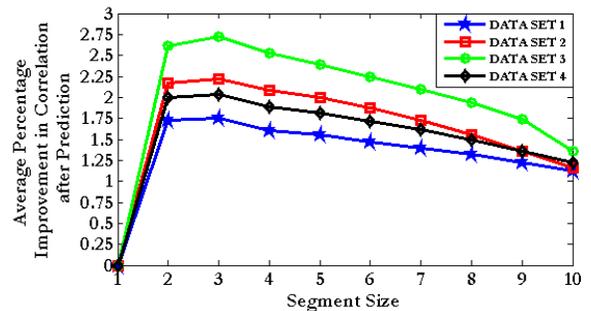

Fig. 9. Effect of changing segment size on improvement in prediction

From the graph, the optimum segment size is 3. As the segment size is increased, the average improvement in the correlation after prediction reduces. This is consistently true for many data sets including the experimentally obtained data.

## B. Prediction at different time lags

The phase screen number to be predicted was changed from 1 to 10 which correspond to a time lag from 0.7-7ms. The effect on the prediction accuracy with changing time lag is shown in Fig. 10. The percentage improvement in the extent of correlation after prediction is above 85% for time lags less than 2.5ms. Improvement goes below 70% for phase screen number greater than 9 (6.1ms).

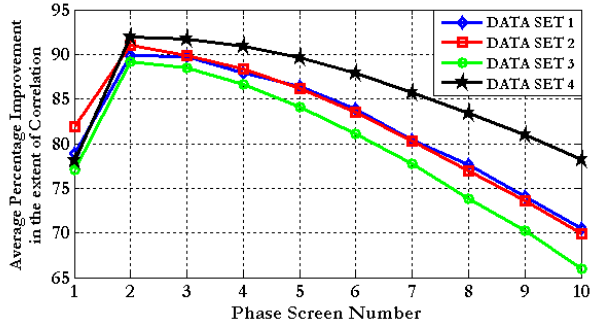

Fig. 10. Percentage improvement at different time lags

## IV. CONCLUSIONS AND DISCUSSION

An attempt has been made to predict Zernike moments corresponding to the incoming wave-fronts distorted due to atmospheric turbulence using data mining. The proposed modal estimation algorithm was implemented on the experimental and simulated data sets. It is shown with simulations as well as experiment that the errors induced by the servo lag delays in adaptive optics imaging systems can be minimized by using this technique. Nearly 3% improvement is observed on an average. The simulation parameters, the segment size and the servo lag timescales were optimized. Segmentation size of 3 was found to be optimum. Large segment size leads to poor prediction. Prediction becomes difficult and inaccurate if the servo lag becomes greater than 6ms. Higher order polynomial interpolation led to worse prediction.